\documentclass[fleqn,twoside]{article}

\usepackage[centertags]{amsmath}
\allowdisplaybreaks[1]

\usepackage{amsbsy}
\usepackage{amsfonts}
\usepackage{amssymb}

\usepackage[dvips]{graphicx}

\usepackage{espcrc2}

\usepackage{url}
\usepackage[dvips]{hyperref}
\usepackage{cite}

\title{Phenomenology of Absolute Neutrino Masses}

\author{Carlo Giunti\address{INFN,
Sezione di Torino,
and
Dipartimento di Fisica Teorica,
Universit\`a di Torino,
Via P. Giuria 1, I--10125 Torino, Italy}}

\begin{document}

\begin{abstract}
The phenomenology of absolute neutrino masses is reviewed,
focusing on
tritium $\beta$ decay,
cosmological measurements
and
neutrinoless double-$\beta$ decay.
\textsf{
Talk presented at NOW-2004,
Neutrino Oscillation Workshop, 11--17 September 2004, Conca Specchiulla, Otranto, Italy.
\null\vspace{-0.5cm}\null
}
\end{abstract}

\maketitle

\section{Introduction}
\label{Introduction}

\begin{figure}[htb]
\begin{center}
\begin{tabular}{lr}
\includegraphics*[bb=183 469 425 771, width=0.45\linewidth]{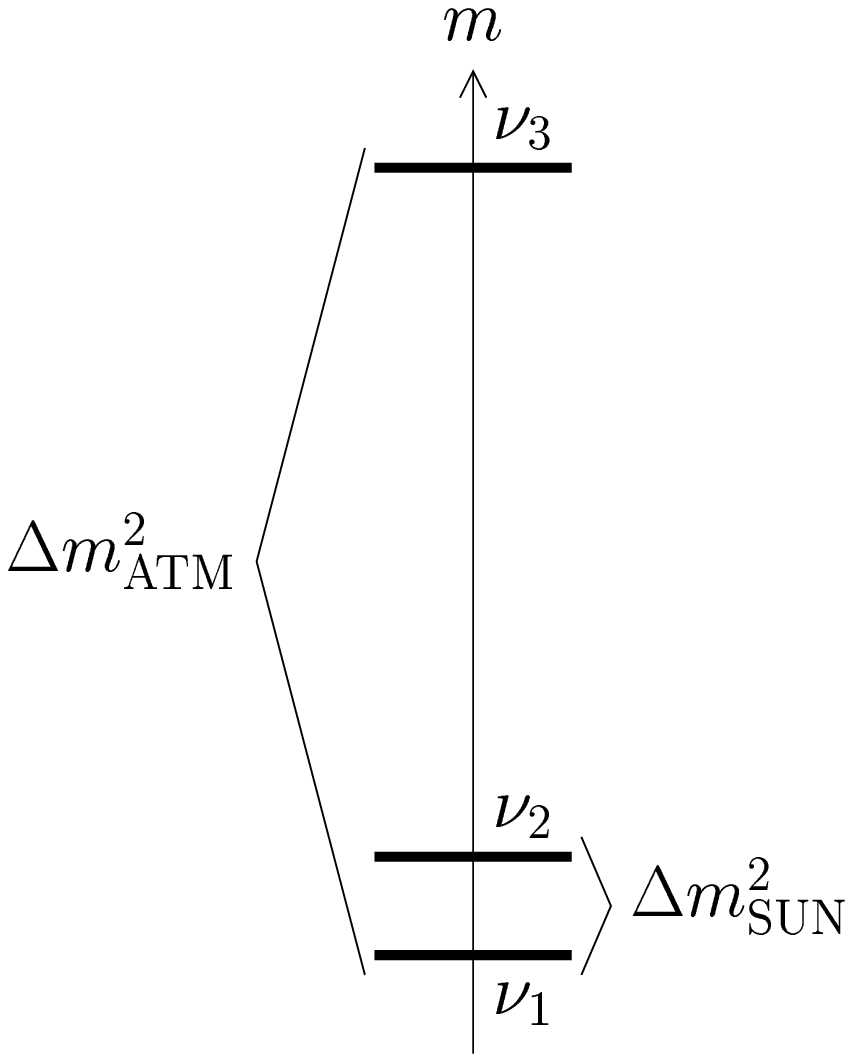}
&
\includegraphics*[bb=186 469 428 771, width=0.45\linewidth]{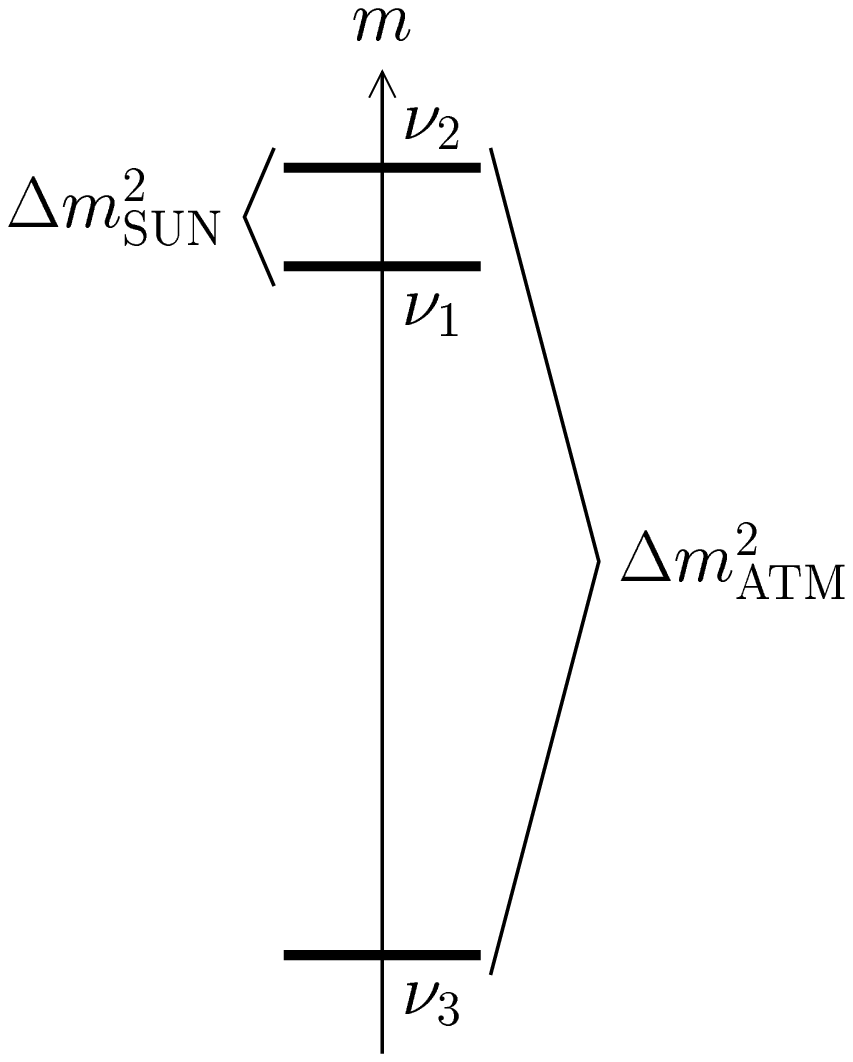}
\\
\textbf{normal}
&
\textbf{inverted}
\end{tabular}
\end{center}
\null\vspace{-1.5cm}\null
\caption{ \label{3nu}
The two three-neutrino schemes allowed by the hierarchy
$\Delta{m}^2_{\text{SUN}} \ll \Delta{m}^2_{\text{ATM}}$.
}
\end{figure}

Solar, atmospheric and long-baseline neutrino oscillation experiments
obtained convincing evidence that neutrinos are massive and mixed particles.
Since on one hand solar and KamLAND experiments
and on the other hand atmospheric and K2K experiments
are sensitive to oscillations generated by
different orders of magnitude of differences of
squared neutrino masses,
at least two independent
$\Delta{m}^2_{\mathrm{SUN}} \ll \Delta{m}^2_{\mathrm{ATM}}$
are needed in order to explain the data.
This requirement is satisfied by the simplest
three-neutrino mixing scheme in which the left-handed components
$\nu_{\alpha L}$
of flavor neutrinos
($\alpha=e,\mu,\tau$)
are linear combinations of
the left-handed components
$\nu_{k L}$
of massive neutrinos
($k=1,2,3$):
$
\nu_{\alpha L} = \sum_{k} U_{\alpha k} \, \nu_{kL}
$,
where $U$ is the unitary mixing matrix
(see, for example, Ref.\cite{BGG-review-98,Bilenky:2002aw,hep-ph/0310238}).
In such framework neutrino oscillations depend on two independent
squared-mass differences
$\Delta{m}^2_{21}$
and
$\Delta{m}^2_{31}$,
that we associate by convention to the
squared-mass differences that are effective in
solar and atmospheric neutrino oscillations,
respectively:
$\Delta{m}^2_{21}=\Delta{m}^2_{\mathrm{SUN}}$
and
$|\Delta{m}^2_{31}|=\Delta{m}^2_{\mathrm{ATM}}$.
The absolute value of $\Delta{m}^2_{31}$ is needed because two
type of schemes,
shown in Fig.\ref{3nu}, are allowed by the observed hierarchy
$\Delta{m}^2_{\mathrm{SUN}} \ll \Delta{m}^2_{\mathrm{ATM}}$.
In the normal scheme, which is so-called because it allows a mass hierarchy
$m_1 \ll m_2 \ll m_3$,
$\Delta{m}^2_{31}$ is positive,
whereas in the inverted scheme
$\Delta{m}^2_{31}$ is negative.

A global fit of the data \cite{Fogli:2004as} gives the best-fits
and
$3\sigma$
ranges for the three-neutrino oscillation parameters listed in Tab.\ref{fit}.
The mixing angles
$\vartheta_{12}$,
$\vartheta_{13}$,
$\vartheta_{23}$
belong to the standard parameterization of the mixing matrix
\cite{Eidelman:2004wy},
in which with good approximation
$\vartheta_{12}$
is the solar mixing angle,
$\vartheta_{23}$
is the atmospheric mixing angle,
and
$\vartheta_{13}$
is the CHOOZ mixing angle
\cite{Bilenky:1998tw,hep-ph/0212142}.
In Tab.\ref{fit} we give only the measured values of
$\vartheta_{12}$
and
$\vartheta_{13}$
obtained in the global fit of Ref.\cite{Fogli:2004as},
which are sufficient for the following discussion
on the phenomenology of absolute neutrino masses.

\begin{table}[bt]
\caption{ \label{fit} Best-fit
and
$3\sigma$
range for the three-neutrino oscillation parameters
obtained in the global fit of Ref.\cite{Fogli:2004as}.
}
\begin{center}
\begin{tabular}{|c|c|}
\hline
Parameter & \begin{tabular}{c} Best-Fit \\ $3\sigma$ Range \end{tabular}
\\
\hline
$\Delta{m}^2_{21}$
&
\begin{tabular}{c} 
$8.3 \times 10^{-5} \, \mathrm{eV}^2$
\\
$7.4 \times 10^{-5} \, - \, 9.3 \times 10^{-5} \, \mathrm{eV}^2$
\end{tabular}
\\
\hline
$\sin^2\vartheta_{12}$
&
\begin{tabular}{c} 
$0.28$
\\
$0.22 \, - \, 0.37$
\end{tabular}
\\
\hline
$|\Delta{m}^2_{31}|$
&
\begin{tabular}{c} 
$2.4 \times 10^{-3} \, \mathrm{eV}^2$
\\
$1.8 \times 10^{-3} \, - \, 3.2 \times 10^{-3} \, \mathrm{eV}^2$
\end{tabular}
\\
\hline
$\sin^2\vartheta_{13}$
&
\begin{tabular}{c} 
$0.01$
\\
$0 \, - \, 0.05$
\end{tabular}
\\
\hline
\end{tabular}
\end{center}
\end{table}

\begin{figure*}[bt]
\begin{minipage}[t]{0.47\textwidth}
\begin{center}
\includegraphics*[bb=120 427 463 750, width=0.95\textwidth]{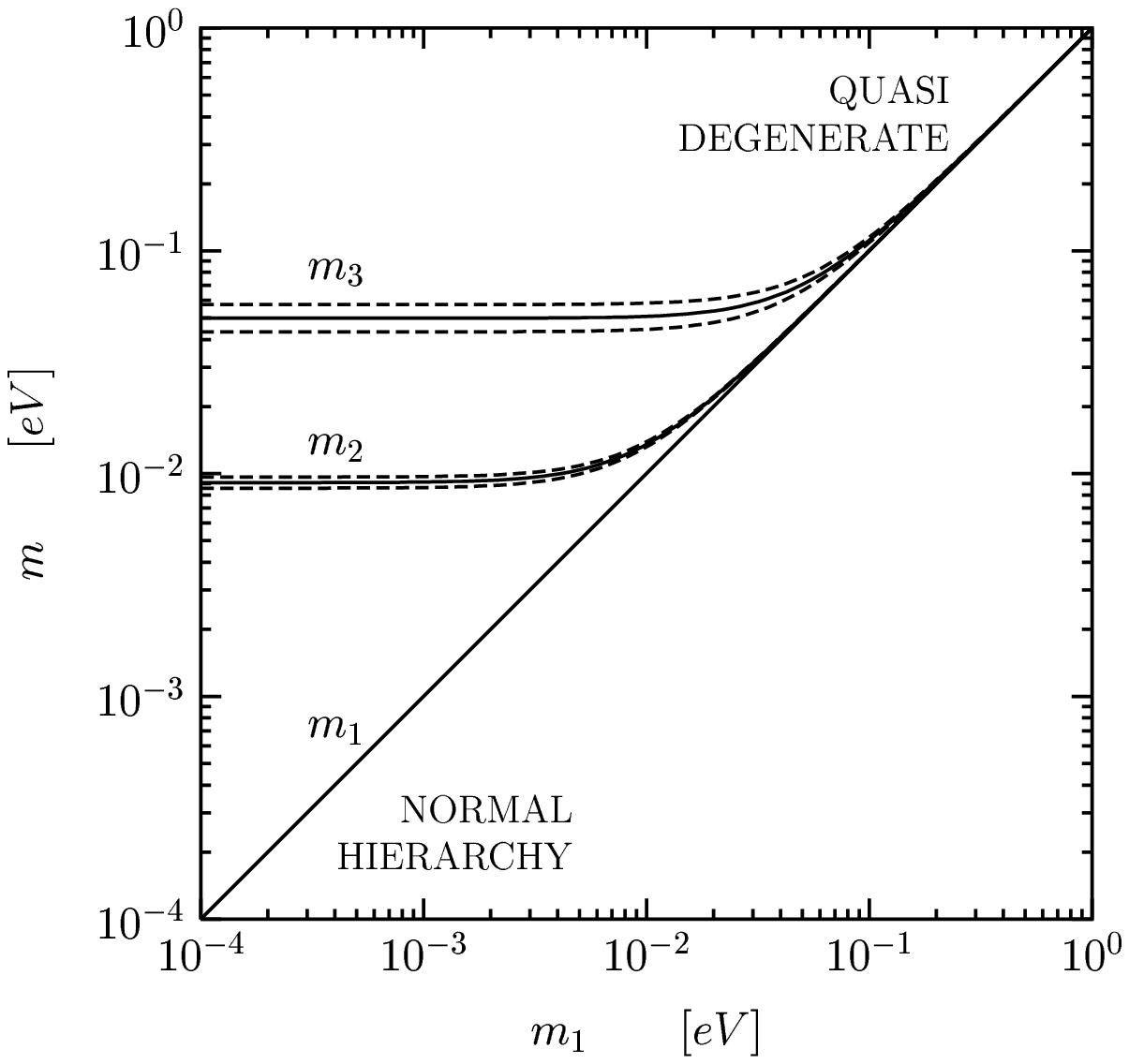}
\end{center}
\end{minipage}
\hfill
\begin{minipage}[t]{0.47\textwidth}
\begin{center}
\includegraphics*[bb=120 427 463 750, width=0.95\textwidth]{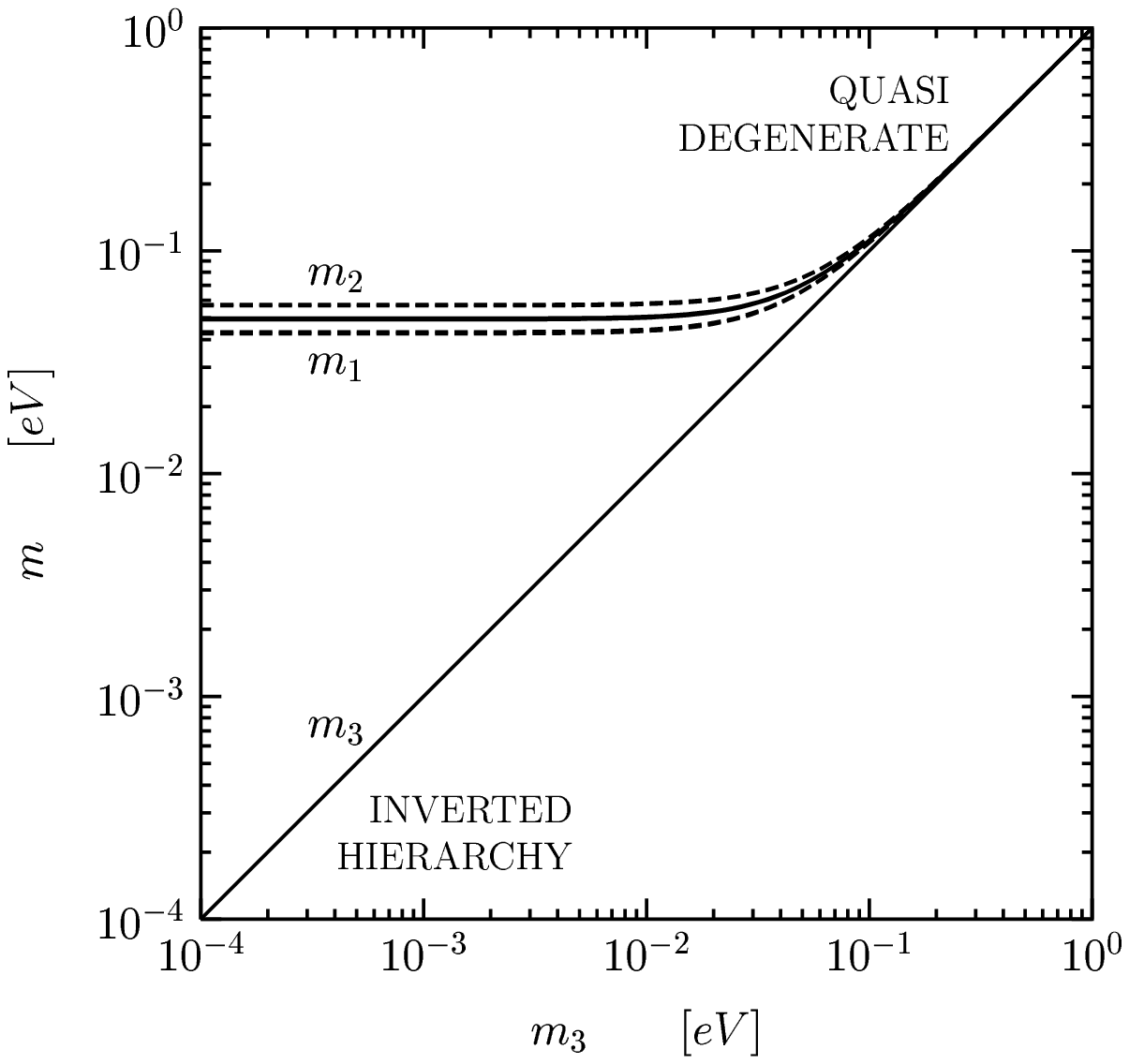}
\end{center}
\end{minipage}
\null\vspace{-1cm}\null
\caption{ \label{3ma}
Values of neutrino masses as functions
of the lightest mass
$m_1$ in the normal scheme and $m_3$ in the inverted scheme.
Solid lines correspond to the best-fit in Tab.~\ref{fit}.
Dashed lines enclose $3\sigma$ ranges.
}
\end{figure*}

Neutrino oscillations depend on the difference of neutrino masses,
not on their absolute value.
As we will see in the following,
other experiments
are able to give information on the absolute value of neutrino masses.
Figure~\ref{3ma} shows the values of
the neutrino masses
obtained from
$\Delta{m}^2_{21}$ and $|\Delta{m}^2_{31}|$ in Tab.\ref{fit}
as functions
of the unknown lightest mass,
which is $m_1$ in the normal scheme
and
$m_3$ in the inverted scheme.
As shown in the figure,
the case $m_3 \ll m_1 \lesssim m_2$
is conventionally called ``inverted hierarchy'',
whereas
in both normal and inverted schemes we have quasi-degeneracy of
neutrino masses for
$
m_1 \simeq m_2 \simeq m_3
\gg
\sqrt{\Delta{m}^2_{\mathrm{ATM}}} \simeq 5 \times 10^{-2} \, \mathrm{eV}
$.

In the following we review the phenomenology
of absolute neutrino masses in
tritium $\beta$ decay (Section~\ref{Tritium}),
cosmological measurements (Section~\ref{Cosmological})
and
neutrinoless double-$\beta$ decay (Section~\ref{Neutrinoless}).

\section{Tritium $\beta$ Decay}
\label{Tritium}

The measurement of the electron spectrum
in $\beta$ decays provides a robust direct determination
of the values of neutrino masses.
In practice the most sensitive experiments
use tritium $\beta$ decay,
because it is a super-allowed transition with a low $Q$-value.
Information on neutrino masses is obtained by measuring the
Kurie function $K(T)$ given by
\cite{Shrock:1980vy,McKellar:1980cn,Kobzarev:1980nk}
$$
K^2(T)
=
( Q - T )
\sum_k
|U_{ek}|^2
\sqrt{ (Q-T)^2 - m_k^2 }
\,,
$$
where $T$ is the electron kinetic energy.
The effect of neutrino masses
can be observed
near the end point of the electron spectrum
where
$Q-T \sim m_k$.
A low $Q$-value is important because
the relative number of events occurring in an interval of energy
$\Delta{T}$
below the end-point is $\propto(T/Q)^3$.

Since the present experiments do not see any effect due to neutrino masses,
it is possible to approximate
$m_k \ll Q-T$
and obtain
\begin{equation}
K^2(T)
\simeq
( Q - T )
\sqrt{ (Q-T)^2 - m_{\beta}^2 }
\,,
\label{g029}
\end{equation}
which is a function of only one parameter,
the effective neutrino mass
\cite{Shrock:1980vy,McKellar:1980cn,Kobzarev:1980nk,Holzschuh:1992xy,Weinheimer:1999tn,Vissani:2000ci,hep-ph/0211341}
\begin{equation}
m_{\beta}^2 = \sum_k |U_{ek}|^2 m_k^2
\,.
\label{g030}
\end{equation}

The current best upper bounds on $m_{\beta}$ are given by the
Mainz and Troitsk experiments
(see Ref.\cite{hep-ex/0210050}),
which obtained the same value
\begin{equation}
m_\beta < 2.2 \, \mathrm{eV}
\qquad
\text{(95\% CL)}
\,.
\label{g032}
\end{equation}
In the future the KATRIN experiment \cite{hep-ex/0309007}
will reach a sensitivity of about $0.2 \, \mathrm{eV}$.

\begin{figure*}[tb]
\begin{minipage}[t]{0.47\textwidth}
\begin{center}
\includegraphics*[bb=102 430 445 754, width=0.95\textwidth]{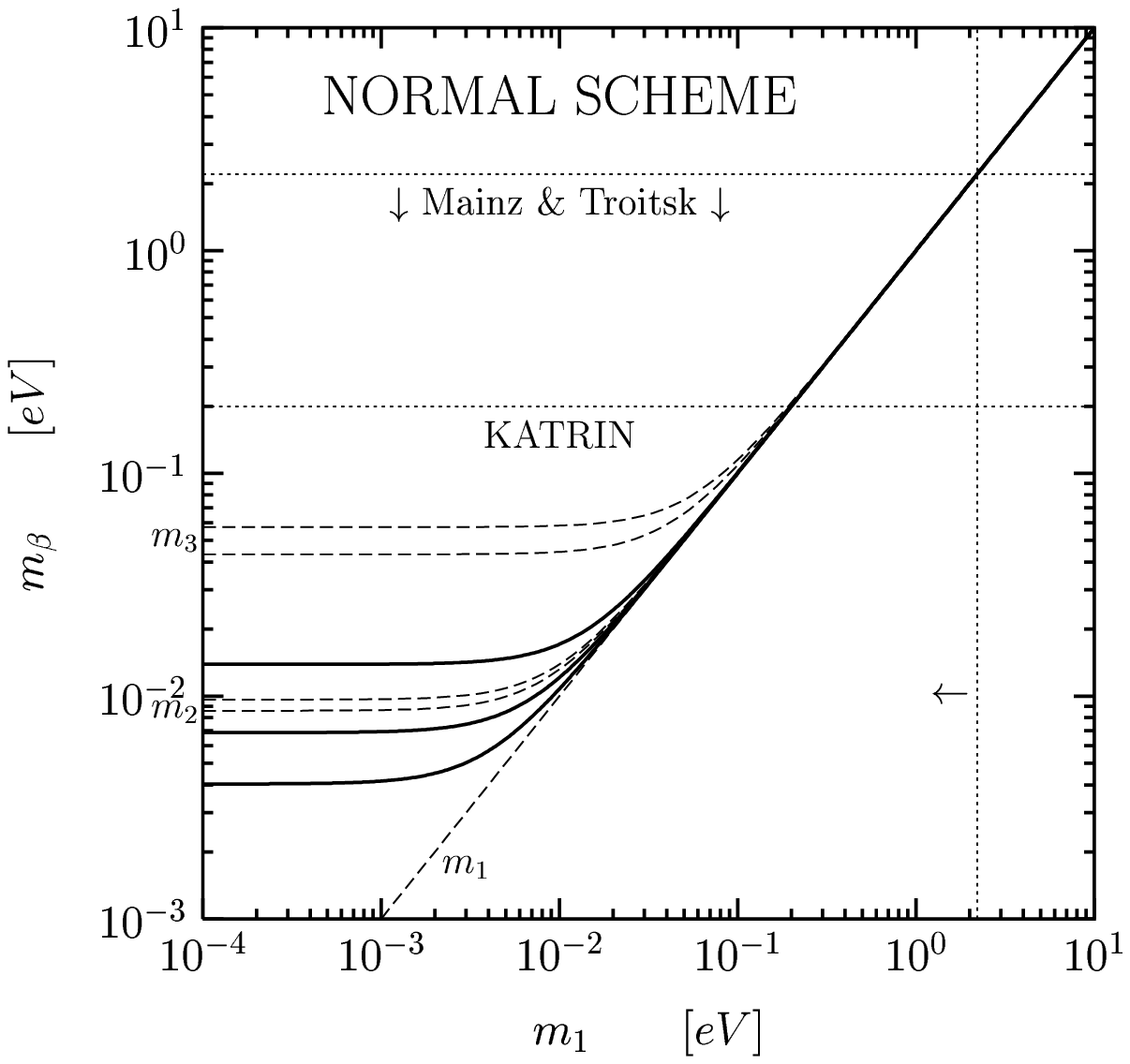}
\end{center}
\end{minipage}
\hfill
\begin{minipage}[t]{0.47\textwidth}
\begin{center}
\includegraphics*[bb=102 430 445 754, width=0.95\textwidth]{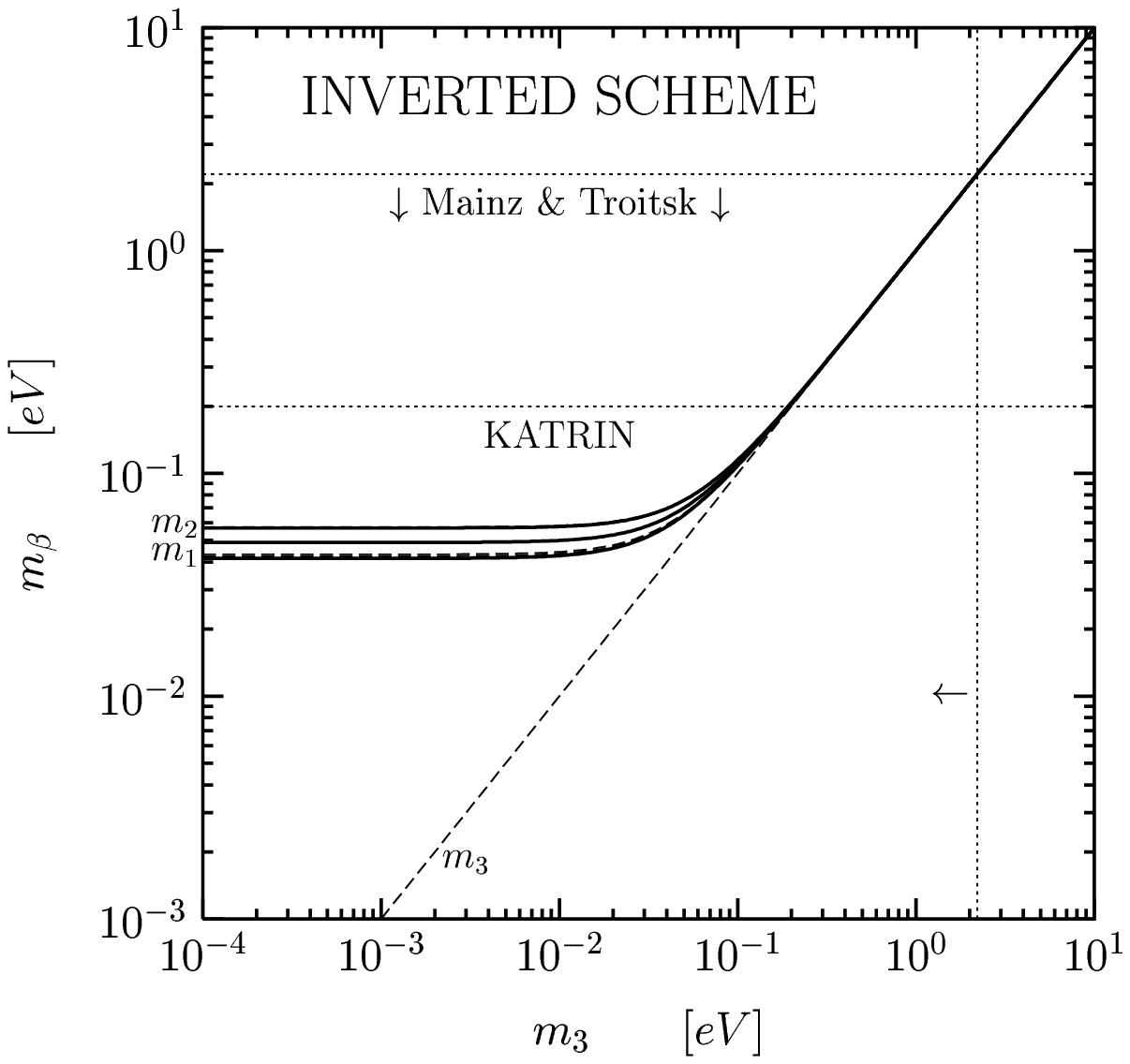}
\end{center}
\end{minipage}
\null\vspace{-1cm}\null
\caption{ \label{mb}
Effective neutrino mass $m_\beta$
in tritium $\beta$-decay experiments as a function
of the lightest mass $m_1$ in the normal scheme and $m_3$ in the inverted scheme.
Middle solid lines correspond to the best-fit in Tab.~\ref{fit}.
Extreme solid lines enclose $3\sigma$ ranges.
Dashed lines delimit $3\sigma$ ranges of individual masses.
}
\end{figure*}

In the standard parameterization of the mixing matrix we have
($c_{ij} \equiv \cos\vartheta_{ij}$
and
$s_{ij} \equiv \sin\vartheta_{ij}$)
\begin{equation}
m_{\beta}^2
=
c_{12}^2 \, c_{13}^2 \, m_1^2
+
s_{12}^2 \, c_{13}^2 \, m_2^2
+
s_{13}^2 \, m_3^2
\,.
\label{g031}
\end{equation}
Since the values of
$\Delta{m}^2_{21}$,
$|\Delta{m}^2_{31}|$,
$\vartheta_{12}$
and
$\vartheta_{13}$
are determined by neutrino oscillation experiments,
there is only one unknown quantity in Eq.(\ref{g031}),
which corresponds to the absolute scale of neutrino masses.
Figure~\ref{mb} shows the value of $m_{\beta}$
as a function
of the unknown lightest mass,
which is $m_1$ in the normal scheme
and
$m_3$ in the inverted scheme,
using the values of the oscillation parameters
in Tab.\ref{fit}.
The middle solid lines correspond to the best fit
and the extreme solid lines delimit the $3\sigma$ allowed range.
We have also shown with dashed lines
the $3\sigma$ ranges of the neutrino masses
(same as in Fig.\ref{3ma}),
which help to understand their contribution to
$m_{\beta}$.
One can see that in the case of a normal mass hierarchy
(normal scheme with
$m_1 \ll m_2 \ll m_3$)
the main contribution to $m_{\beta}$
can be due to $m_2$ or $m_3$ or both,
because the upper limit for
$m_{\beta}$
is larger than the upper limit for
$m_2$.
In the case of an inverted mass hierarchy
(inverted scheme with
$m_3 \ll m_1 \lesssim m_2$)
$m_{\beta}$
has practically the same value as $m_1$ and $m_2$.

Figure~\ref{mb}
shows that the Mainz and Troitsk experiments
and the future KATRIN experiment
give information on the absolute values of neutrino masses
in the quasi-degenerate region
in both normal and inverted schemes.
In the far future the inverted scheme could be excluded
if experiments with a sensitivity of
about
$4 \times 10^{-2} \, \mathrm{eV}$
will not find any effect of neutrino masses.

\section{Cosmological Measurements}
\label{Cosmological}

If neutrinos have a mass of the order of 1 eV
they constitute a so-called ``hot dark matter'',
which suppresses the power spectrum of density fluctuations
in the early universe at ``small'' scales of the order of
1$-$10 Mpc
(see Ref.\cite{Hu:1998mj}).
The suppression depends on
the sum of neutrino masses
$ \sum_k m_k $.

\begin{figure*}[tb]
\begin{minipage}[t]{0.47\textwidth}
\begin{center}
\includegraphics*[bb=102 430 437 748, width=0.95\textwidth]{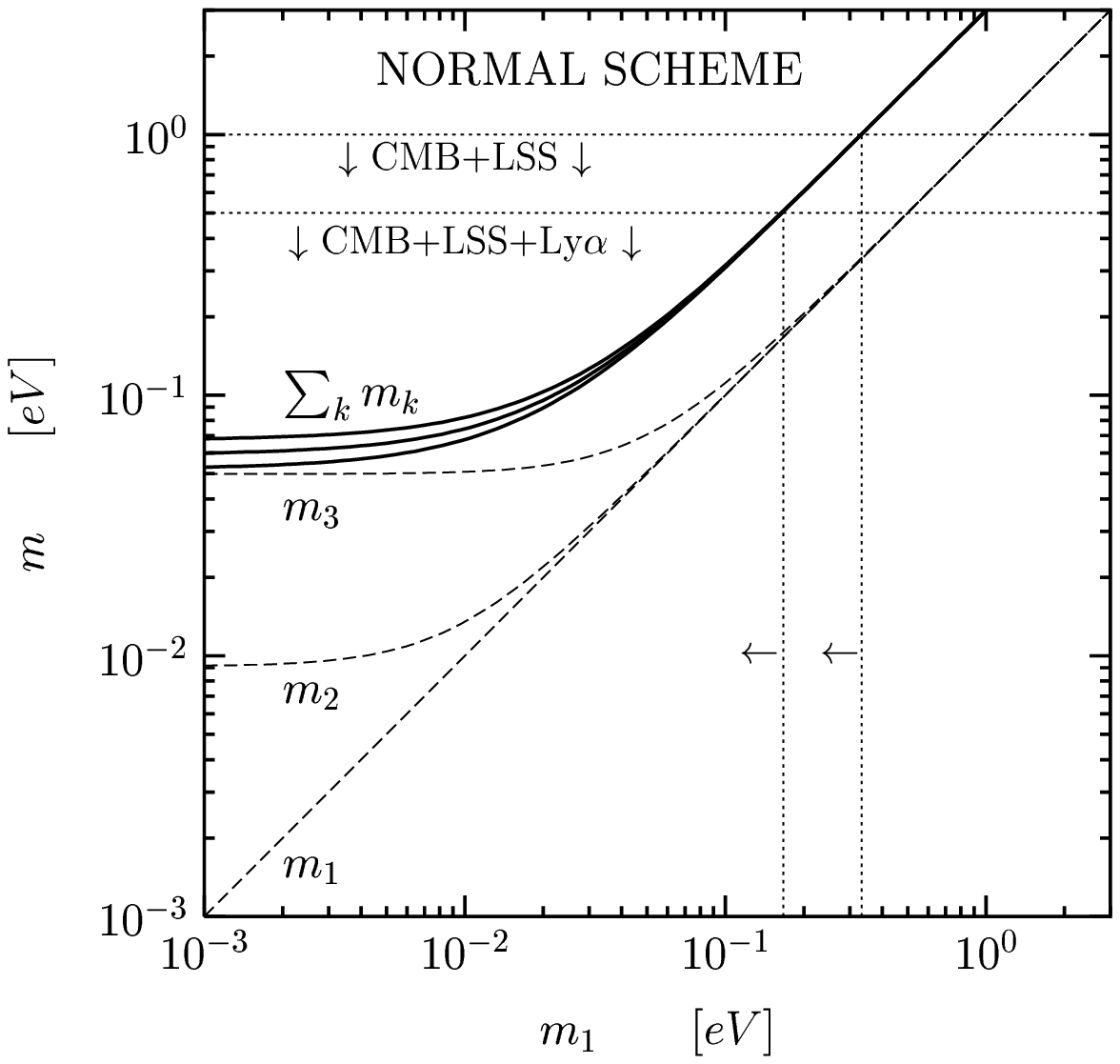}
\end{center}
\end{minipage}
\hfill
\begin{minipage}[t]{0.47\textwidth}
\begin{center}
\includegraphics*[bb=102 430 437 748, width=0.95\textwidth]{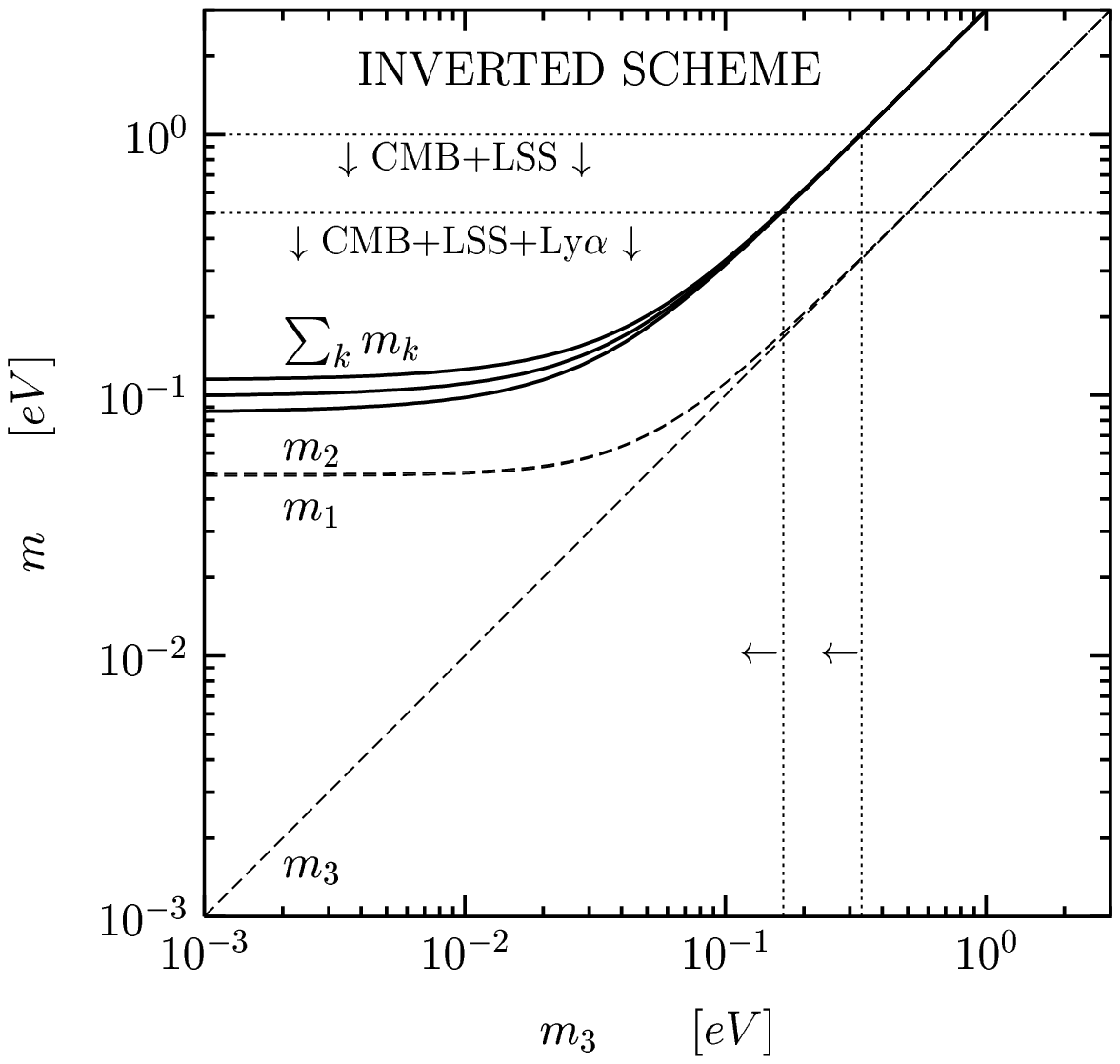}
\end{center}
\end{minipage}
\null\vspace{-1cm}\null
\caption{ \label{cosmo}
Sum of neutrino masses
as a function
of the lightest mass $m_1$ in the normal scheme and $m_3$ in the inverted scheme.
Middle solid lines correspond to the best-fit in Tab.~\ref{fit}.
Extreme solid lines enclose $3\sigma$ ranges.
Dashed lines show the best-fit values of individual masses.
}
\end{figure*}

Recent high precision measurements of
density fluctuations
in the Cosmic Microwave Background
(WMAP)
and
in the Large Scale Structure of galaxies
(2dFGRS, SDSS),
combined with other cosmological data
have allowed to put stringent upper limits on
$ \sum_k m_k $
\cite{Spergel:2003cb,astro-ph/0303076,astro-ph/0303089,astro-ph/0310723,astro-ph/0406594,Seljak:2004xh,Fogli:2004as}.
However,
different authors have obtained significantly different
upper bounds mainly because of the different sets of data considered.
The most crucial data are the so-called Lyman-$\alpha$ forests
which are constituted by absorption lines in the spectra of high-redshift quasars
due to intergalactic hydrogen clouds.
Since these clouds have dimensions of the order of
1$-$10 Mpc,
the Lyman-$\alpha$ data are crucial in order to push the
upper bound on $ \sum_k m_k $ below 1 eV.
Unfortunately the interpretation of
Lyman-$\alpha$ data
may suffer from large systematic uncertainties.
Summarizing the different limits obtained in
Refs.\cite{Spergel:2003cb,astro-ph/0303076,astro-ph/0303089,astro-ph/0310723,astro-ph/0406594,Seljak:2004xh,Fogli:2004as},
we estimate the approximate $2\sigma$ upper bounds
\begin{equation}
\sum_k m_k \lesssim 0.5 \, \mathrm{eV}
\,,
\qquad
\sum_k m_k \lesssim 1 \, \mathrm{eV}
\,,
\label{c01}
\end{equation}
with and without Lyman-$\alpha$ data, respectively.
From Fig~\ref{cosmo}
one can see that both these limits constrain the neutrino masses
in the quasi-degenerate region,
where the upper bound on each individual mass
is one third of the bound on the sum.
In the future the inverted scheme can be excluded by an upper bound
of about
$8 \times 10^{-2} \, \mathrm{eV}$
on the sum of neutrino masses.

\section{Neutrinoless Double-$\beta$ Decay}
\label{Neutrinoless}

Neutrinoless double-$\beta$ decay is a very important process
because it is not only sensitive to the absolute value of neutrino masses,
but mainly because it is allowed only if neutrinos
are Majorana particles
\cite{Schechter:1982bd,Takasugi:1984xr}.
A positive result in neutrinoless double-$\beta$ decay
would represent a discovery of a new type of particles,
Majorana particles,
a fundamental improvement in our understanding of nature.

Neutrinoless double-$\beta$ decays are processes of type
$
\mathcal{N}(A,Z)
\to
\mathcal{N}(A,Z+2)
+
e^-
+
e^-
$,
in which no neutrino is emitted,
with a change of two units of the total lepton number.
These processes,
forbidden in the Standard Model,
have half-lives
\begin{equation}
T_{1/2}^{0\nu}
=
\left(
G_{0\nu}
\,
|\mathcal{M}_{0\nu}|^2
\,
|m_{\beta\beta}|^2
\right)^{-1}
\,,
\label{d02}
\end{equation}
where $G_{0\nu}$ is the phase-space factor,
$\mathcal{M}_{0\nu}$
is the nuclear matrix element
and
\begin{equation}
m_{\beta\beta}
=
\sum_k
U_{ek}^2 \, m_k
\label{d03}
\end{equation}
is the effective Majorana mass.

A possible indication of neutrinoless double-$\beta$ decay of ${}^{76}\mathrm{Ge}$
with half-life
\begin{equation}
T_{1/2}^{0\nu}({}^{76}\mathrm{Ge})
=
( 0.69 - 4.18 ) \times 10^{25} \, \mathrm{y}
\qquad
(3\sigma)
\label{d05}
\end{equation}
has been found by the authors of Ref.\cite{hep-ph/0404088},
whereas other experiments found only lower bounds.
The most stringent lower bound on $T_{1/2}^{0\nu}({}^{76}\mathrm{Ge})$
has been obtained in the Heidelberg-Moscow experiment \cite{Klapdor-Kleingrothaus:2001yx}:
\begin{equation}
T_{1/2}^{0\nu}({}^{76}\mathrm{Ge})
>
1.9 \times 10^{25} \, \mathrm{y}
\qquad
\text{(90\% CL)}
\,.
\label{d06}
\end{equation}
The IGEX experiment \cite{Aalseth:2002rf}
obtained the comparable limit
$
T_{1/2}^{0\nu}({}^{76}\mathrm{Ge})
>
1.57 \times 10^{25} \, \mathrm{y}
$
(90\% CL).
Hence, the status of the experimental search for neutrinoless double-$\beta$
decays is presently uncertain and new experiments which can check
the indication (\ref{d05}) are needed
(see Ref.\cite{hep-ph/0405078}).

The extraction of the value of $|m_{\beta\beta}|$
from the data
has unfortunately a serious problem due to the
large theoretical uncertainty in the evaluation of the
nuclear matrix element
$\mathcal{M}_{0\nu}$
(see Refs.\cite{Civitarese:2002tu,hep-ph/0405078}).
In the following we will use as ``$3\sigma$'' range for the
nuclear matrix element
$|\mathcal{M}_{0\nu}|$
the interval which covers the results of reliable calculations
listed in Tab.2 of Ref.\cite{hep-ph/0405078}
(other approaches are discussed in Refs.\cite{Rodin:2003eb,hep-ph/0402250,hep-ph/0403167,Fogli:2004as}):
\begin{equation}
0.41
\lesssim
|\mathcal{M}_{0\nu}|
\lesssim
1.24
\,,
\label{d04}
\end{equation}
which corresponds to a ``$3\sigma$'' uncertainty of a factor of 3 for the
determination of $|m_{\beta\beta}|$ from $T_{1/2}^{0\nu}({}^{76}\mathrm{Ge})$.
Using the range (\ref{d04}),
the indication (\ref{d05}) implies
\begin{equation}
0.22 \, \mathrm{eV}
\lesssim
|m_{\beta\beta}|
\lesssim
1.6 \, \mathrm{eV}
\,,
\label{d08}
\end{equation}
and the most stringent upper bound (\ref{d06})
implies
\begin{equation}
|m_{\beta\beta}|
\lesssim
0.32 - 1.0 \, \mathrm{eV}
\,.
\label{d09}
\end{equation}

In the standard parameterization of the mixing matrix
the effective Majorana mass is given by
$$
m_{\beta\beta}
=
c_{12}^2 \, c_{13}^2 \, m_1
+
s_{12}^2 \, c_{13}^2 \, e^{i\alpha_{21}} \, m_2
+
s_{13}^2 \, e^{i\alpha_{31}} \, m_3
\,,
$$
where
$\alpha_{21}$ and $\alpha_{31}$
are unknown Majorana phases
(see, for example, Ref.\cite{BGG-review-98,Bilenky:2002aw,hep-ph/0310238}).

\begin{figure*}[tb]
\begin{minipage}[t]{0.47\textwidth}
\begin{center}
\includegraphics*[bb=120 427 463 750, width=0.95\textwidth]{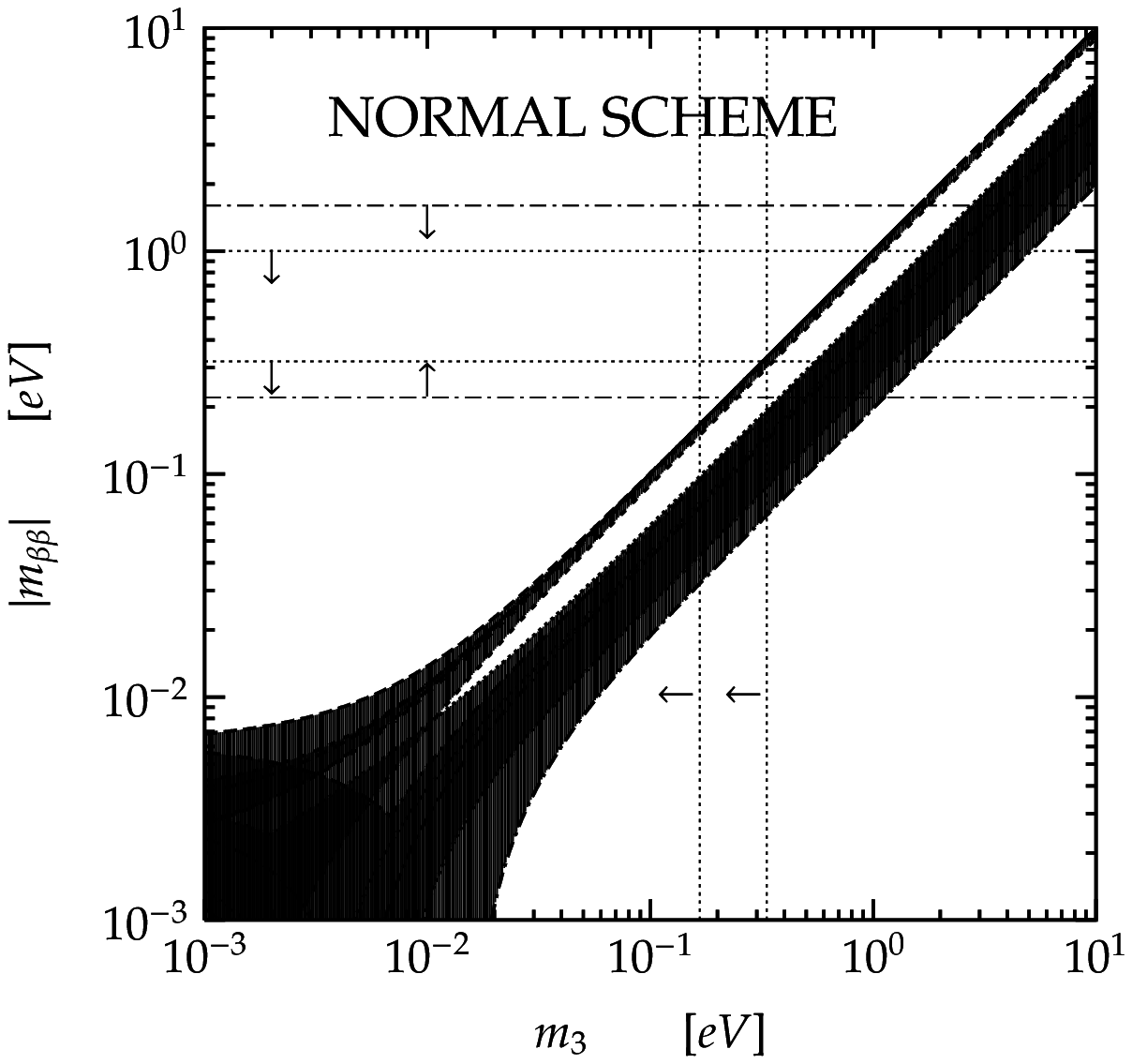}
\end{center}
\end{minipage}
\hfill
\begin{minipage}[t]{0.47\textwidth}
\begin{center}
\includegraphics*[bb=120 427 463 750, width=0.95\textwidth]{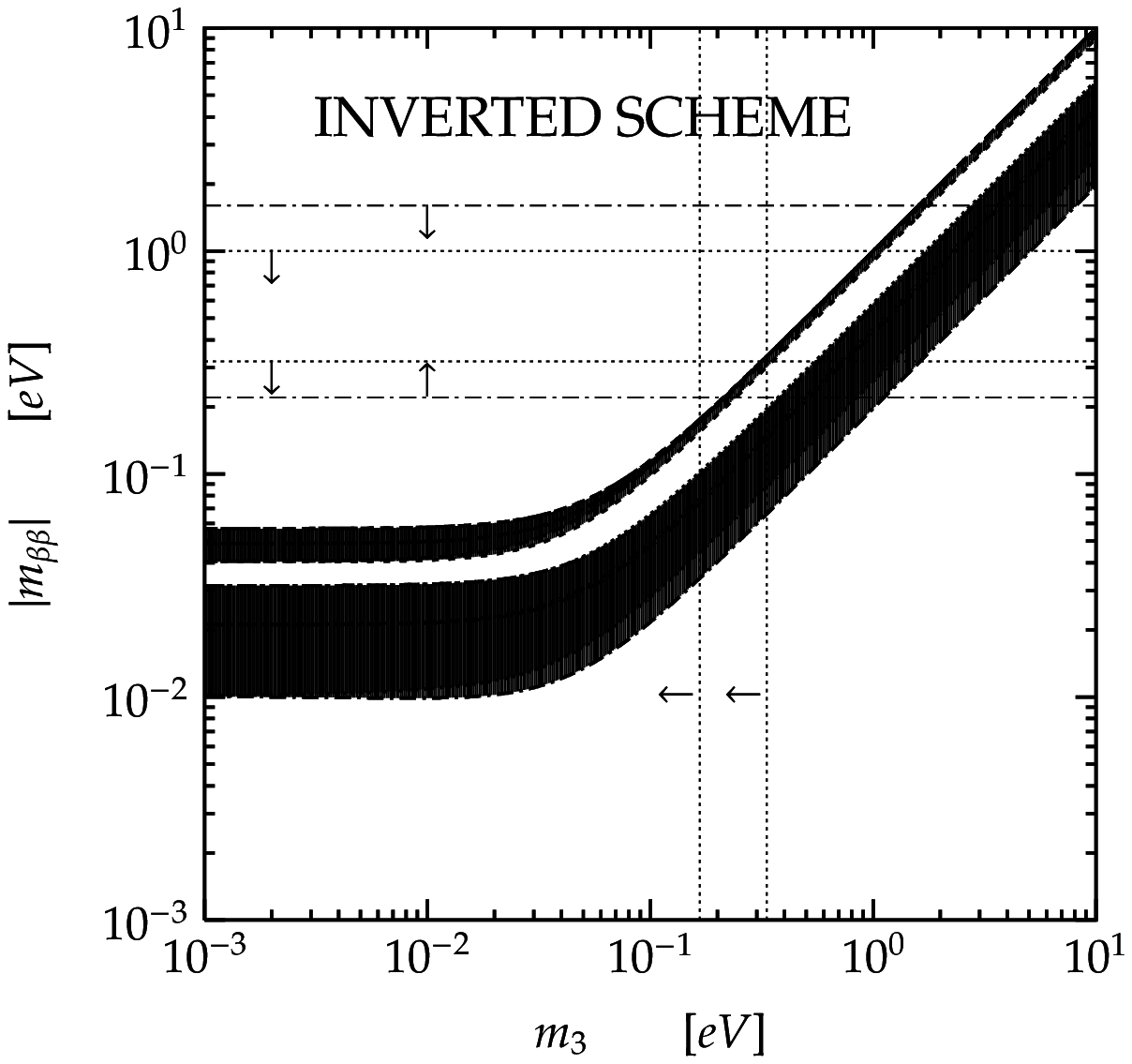}
\end{center}
\end{minipage}
\null\vspace{-1cm}\null
\caption{ \label{db}
Effective Majorana mass $|m_{\beta\beta}|$
in neutrinoless double-$\beta$ decay experiments as a function
of the lightest mass $m_1$ in the normal scheme and $m_3$ in the inverted scheme.
The white areas in the strips need CP violation.
The horizontal dotted lines show the interval (\ref{d09})
of uncertainty of the current experimental upper bound
due to the estimated uncertainty (\ref{d04})
of the value of the nuclear matrix element.
The horizontal dash-dotted lines delimit the range (\ref{d08})
obtained from the indication (\ref{d05}).
The vertical dotted lines correspond to the cosmological upper
bounds on individual neutrino masses in Fig.\ref{cosmo}.
}
\end{figure*}

Figure~\ref{db} shows the allowed range for
$|m_{\beta\beta}|$
obtained with the mixing parameters in Tab.\ref{fit}
(see also
Refs.\cite{Feruglio:2002af-ADD,hep-ph/0304276,hep-ph/0310003,hep-ph/0310238,hep-ph/0402250,Bahcall:2004ip,Petcov-NJP6-109-2004}).
One can see that in the region where the lightest mass is very small
the allowed ranges for $|m_{\beta\beta}|$
in the normal and inverted schemes are dramatically different.
This is due to the fact that
in the normal scheme strong cancellations between the contributions
of $m_2$ and $m_3$ are possible,
whereas in the inverted scheme
the contributions of $m_1$ and $m_2$ cannot cancel because maximal mixing
in the 1$-$2 sector is excluded by solar data
($\vartheta_{12}<\pi/4$ at $5.8\sigma$ \cite{Bahcall:2004ut}).
On the other hand,
there is no difference between the normal and inverted schemes
in the quasi-degenerate region,
which is probed by the present data.
From Fig.\ref{db} one can see that there is a tension
between the indication (\ref{d08})
and the cosmological upper
bound on individual neutrino masses,
especially the one with Lyman-$\alpha$ data.
In the future,
the normal and inverted schemes may be distinguished by reaching a sensitivity
of about $10^{-2} \, \mathrm{eV}$.

\section{Conclusions}
\label{Conclusions}

In conclusion we would only like to emphasize the
fundamental importance of the determination
of the Dirac or Majorana nature of neutrinos
and their absolute mass scale.
Improvements on these topics
should be strongly pursued in future
experimental and theoretical research.

\end{document}